\documentclass[runningheads]{llncs}
\usepackage[T1]{fontenc}
\usepackage{comment}
\usepackage{graphicx}
\usepackage{amsmath}
\usepackage{amssymb}
\usepackage{booktabs}
\usepackage{appendix}
\usepackage{xcolor}
\usepackage{multirow}

\usepackage[misc]{ifsym} 

\begin{document}
\bibliographystyle{splncs04}
\title{Co-Learning Semantic-aware Unsupervised Segmentation for Pathological Image Registration} 
\titlerunning{Collaborative Learning of GIRNet}

\author{
Yang Liu\inst{1} \and
Shi Gu\inst{1,2}$^{(\textrm{\Letter})}$
}

\authorrunning{Y. Liu et al.}

\institute{
University of Electronic Science and Technology of China, Chengdu, China \\
Shenzhen Institute for Advanced Study, UESTC \\
\email{gus@uestc.edu.cn}
}

\maketitle      

\begin{abstract}

The registration of pathological images plays an important role in medical applications. Despite its significance, most researchers in this field primarily focus on the registration of normal tissue into normal tissue. The negative impact of focal tissue, such as the loss of spatial correspondence information and the abnormal distortion of tissue, are rarely considered. In this paper, we propose \textbf{GIRNet}, a novel unsupervised approach for pathological image registration by incorporating segmentation and inpainting through the principles of Generation, Inpainting, and Registration (GIR). The registration, segmentation, and inpainting modules are trained simultaneously in a co-learning manner so that the segmentation of the focal area and the registration of inpainted pairs can improve collaboratively. Overall, the registration of pathological images is achieved in a completely unsupervised learning framework. Experimental results on multiple datasets, including Magnetic Resonance Imaging (MRI) of T1 sequences, demonstrate the efficacy of our proposed method. Our results show that our method can accurately achieve the registration of pathological images and identify lesions even in challenging imaging modalities. Our unsupervised approach offers a promising solution for the efficient and cost-effective registration of pathological images. Our code is available at \url{https://github.com/brain-intelligence-lab/GIRNet}.


\end{abstract}
\section{Introduction}
Image registration has been widely studied in both academia and industry over the past two decades. In general, the goal of deformable image registration is to estimate a suitable nonlinear transformation that overlaps the pair of images with corresponding spatial relationships \cite{beg2005computing,wu2022nodeo}. This goal is usually achieved by minimizing a well-defined similarity score. However, these methods often assume that there is no spatial non-correspondence between the two images. In the field of medical image analysis, this assumption is often not valid, particularly in cases such as pathology image to atlas registration  or pre-operative and post-operative longitudinal registration. Direct registration of pathology images without taking into account the impact of focal tissue can result in missed pixel-level correspondence and large registration errors.

A variety of approaches have been proposed to handle the non-correspondence problem in medical image registration. These methods can be roughly divided into three main categories: \emph{1) Cost function masking.} The authors of \cite{ref_article3,ref_article4} used the segmentation of the non-corresponding regions to mask the image similarity measure in optimization. \emph{2) Converting pathological image to normal appearance}. This class of approaches aims to replace or reconstruct the focal area as normal tissue to guide the registration either through low-rank and sparse image decomposition \cite{ref_article7,ref_article8} or generative models \cite{ref_article6}. \emph{3) Non-correspondence detection via intensity criteria}. This category of methods can be formulated as joint segmentation and registration to detect non-corresponding regions during the registration process \cite{ref_article13,ref_article18}. Although these approaches partially handle the issue of non-correspondence in the registration, they still have some serious shortcomings. The cost function masking and image conversion approaches require ground truth or accurate labels during registration and may decrease the alignment accuracy when the focal area is large. The non-correspondence detection approach, which typically relies on a sophisticated designed loss function, is very sensitive to the dataset \cite{ref_article19} and difficult to find a set of unified parameters.

Therefore, to effectively address the non-correspondence problem in registering pathology images, it is necessary to incorporate both a data-independent segmentation module and a modality-adaptive inpainting module into the registration pipeline. 
To bridge this gap, we introduce the semantic information of the category based on \cite{ref_article24,ref_article25}. It employs the non-correspondence in registration to achieve accurate segmentation of the lesion region and uses the segmented mask to reconstruct the lesion area and guide the registration. In this paper, we address the challenge of large alignment errors due to the loss of spatial correspondence in processing pathological images. To overcome this challenge, we propose a tri-net collaborative learning framework that simultaneously updates the registration, segmentation, and inpainting networks. The segmentation network minimizes the mutual information between the lesion and normal tissue based on the semantic information introduced by the registration network, allowing for accurate segmentation of regions with missing spatial correspondence. The registration network, in turn, weakens the adverse effects of the lesions based on the mask generated by the segmentation network. To the best of our knowledge, this is the first work to apply an unsupervised segmentation method based on minimal mutual information (\textbf{MMI}) to pathological image registration, with simultaneous training of segmentation and registration. Our work makes the following key contributions.\begin{itemize}
\item We propose a collaborative learning method for the simultaneous optimization of registration, segmentation, and inpainting networks.
\item
We show the effectiveness of using mutual information minimization in an unsupervised manner for pathological image segmentation and registration by incorporating semantic information through the registration process.
\item
We perform a series of experiments to validate our method's superiority in accurately finding lesions and effectively registering pathological images.
\end{itemize}

\section{Method}
\begin{figure}[!t]
\includegraphics[width=\textwidth]{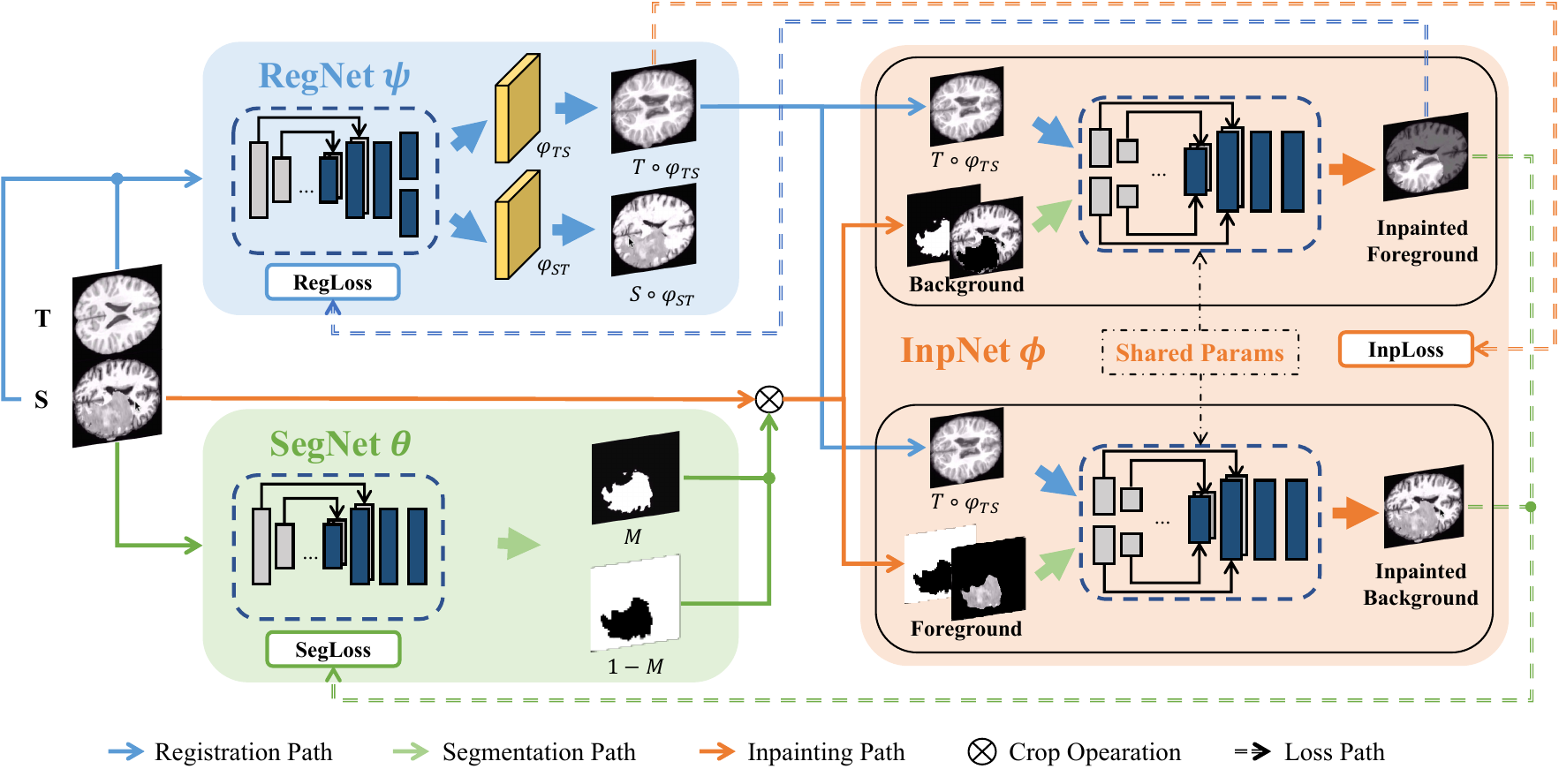}
\caption{The proposed tri-modules collaborative learning framework for medical image analysis includes RegNet, SegNet, and InpNet to achieve accurate image registration and segmentation through the optimization of semantic-informed mutual information.}
\label{fig2}
\end{figure}

Our proposed framework (Figure \ref{fig2}) involves three modules:  a register denoted by $\psi$, a segmenter denoted by $\theta$, and an inpainter denoted by $\phi$. The three modules are trained in a co-learning manner to enable the registration aware of semantic information. Importantly, our proposed training procedure is fully unsupervised which does not require any labeled data for training the network.

\subsection{Collaborative optimization}
The most critical problem in pathological image registration is identifying and dealing with the lesion area. If we naively register a source pathological image $S$ to a template $T$ without caring about the lesion boundary, the deformation field near the boundary would be uncontrollable becuase a healthy template does not have a lesion. A possible approach here is to initialize an inflating boundary containing the lesion area, followed by calculating the registration loss either outside of the boundary only or based on a modified $S$ that is inpainted within the given boundary. However, the registration error has no sensitivity to the location of the inflated boundary as long as it is larger than the real one. On the other hand, if we compared the inpainted image and the pathological image $S$ within the boundary only, we can notice that their dissimilarity increases when the boundary shrinks as the inpainting algorithm only generates healthy parts. This mechanism can then induce a segmentation module that segments the lesion as the foreground and the remaining as the background, which iteratively serves as the input mask for the inpainting module. Further, as the registration loss is calculated based on the registered inpainted image and the target image, the registration provides a regularization for the inpainting module such that the inpainting is specialized to facilitate the registration.

Specially for the input and output of the three modules, RegNet takes images $S$ and $T$ as input and generates the deformation field from $S$ to $T$ and $T$ to $S$ as $\varphi_{ST}$ and $\varphi_{TS}$ respectively. InpNet takes the background (foreground) cropped by SegNet and image $T\circ \varphi_{TS}$ warped by RegNet as input and outputs foreground (background) with a normal appearance. SegNet takes the pathology image $S$ as input and employs the normal foreground and background inpainted by InpNet to segment the lesion region based on MMI. SegNet and InpNet are actually in an adversarial relationship.
Through this joint optimization approach, the three networks collectively work to achieve registration and segmentation of pathological images under entirely unsupervised conditions, without being limited by the specific network structure. For the sake of simplicity, we employ a Unet-like \cite{ref_article30} basic structure without any normalization layer.\\

\subsection{Network Modules}

\subsubsection{RegNet.}
The primary objective of registration is to generate a deformation field that minimizes the dissimilarity between the source image (S) and the template image (T). The deformation is usually required to satisfy constraints like smoothness and even diffeomorphism.
In terms of pathological image registration, the deformation field is only valid 
off the lesion area. Thus the registration loss should be calculated on the normal area only. Suppose that the lesion area is already obtained as $\theta(S)$ and inpainted with normal tissue, the registration loss can then be formulated as
\begin{equation}
    \label{eq7}
    \begin{split}
         \mathcal{L}_{reg} = \min\limits_{\psi} \Bigl\{
         & \mathcal{L}_{sym} (\phi(S \cdot \overline{\theta(S)}|T\circ \varphi_{TS})\circ \varphi_{ST}, T) \\
        &+\mathcal{L}_{sym}(T\circ \varphi_{TS}, \phi(S \cdot \overline{\theta(S)}|T\circ \varphi_{TS}))
        \Bigl\}
    \end{split}
\end{equation}
where $\varphi_{ST}=\psi(S,T)$, $\varphi_{TS}=\psi(T,S)$ are the deformation fields that warp $S\rightarrow T$ and $T\rightarrow S$ respectively. The symbol $\cdot$ denotes element-wise multiplication and $\mathcal{L}_{sym}$ is the registration loss of SymNet \cite{ref_article31} that balance the losses of orientation consistency, regularization and magnitude.

\subsubsection{SegNet.}
Minimal Mutual Information (MMI) is a typically used unsupervised segmentation method that distinguishes foreground from background. However, for a pathological image, the lesion regions often have a similar intensity to normal tissues near the boundary, which prevents the MMI from accurate segmentation without the semantic information. To address this limitation, we warp a healthy image $T$ onto a pathology image $S$ using a deformation field $\varphi_{TS}=\psi(T,S)$. This process maximizes the mutual information between corresponding regions of the two images and minimizes that of non-corresponding regions, thereby facilitating accessible lesion segmentation with MMI. Let $\Omega \in \mathbb{R^d}$ denote the image domain, $F_{\theta}=\Omega \circ M$ and $B_{\theta}=\Omega \circ \overline{M}$ denote the foreground and background, where $\overline{M}=1-M, M \in \left \{0, 1\right \}$. Regarding a pathological image $S$, when the background (normal) is given, the inpainted forground (normal) will be different from the true foreground (lesion). When the foreground (lesion) is given, the inpainted background will remain the same as the background (normal). Thus we can formulate the adversarial loss of unsupervised segmentation as
\begin{equation}
    \begin{split}
    \label{eq6}
    \mathcal{L}_{seg} = \max\limits_{\theta}\min\limits_{\phi} &\left\{\frac{\mathbb{E}\{ \theta(S)\cdot \mathcal{D}[S,  \phi(S \cdot \overline{\theta(S)}|T\circ \varphi_{TS})]\}}{\mathbb{E} \left \|\theta(S) \right \|} \right.\\ & \left. -\frac{\mathbb{E}\{ \overline{\theta(S)}\cdot \mathcal{D}[S,  \phi(S \cdot \theta(S)| T\circ \varphi_{TS})]\}}{\mathbb{E} \left \|\overline{\theta(S)}\right\|}\right\} ,
    \end{split}
\end{equation}
where $\mathcal{D}$ is the distance function given by localized normalized cross-correlation ($\mathbf{LNCC}$) \cite{ref_article29}. Appendix A provides a detailed derivation.

\subsubsection{InpNet.}
Let $M$ denote the mask and $\varphi_{TS}$ denote the deformation field from $T$ to $S$. To handle the potential domain differences between the masked image $S\circ M$ and the aligned image $T\circ \varphi_{TS}$, InpNet employs two encoders. The adversarial loss function of InpNet is represented as $\mathcal{L}_{MI}$. To incorporate semantic information, we include an additional similarity term $\mathcal{L}_{sim}$ that prevents InpNet from focusing too heavily on the foreground (lesion) and encourages it to produce healthy tissue. The proposed loss function $\mathcal{L}_{inp}$ is then formulated as the combination of mutual information loss defined through the normalized correlation coefficient (NCC) and similarity loss through the mean squared error (MSE):
\begin{equation}
\begin{split}
        \label{eq8}
        \mathcal{L}_{inp} = &\mathcal{L}_{MI} + \lambda \mathcal{L}_{sim},
\end{split}
\end{equation}
with
\begin{equation}
\begin{split}
        \mathcal{L}_{MI} = &\mathcal{L}_{NCC}(S, \phi(S \cdot \overline{\theta(S)}|T \circ \varphi_{TS})) + \mathcal{L}_{NCC}(S, \phi(S \cdot \theta(S)|T \circ \varphi_{TS})), \\
        \mathcal{L}_{sim} = & \mathcal{L}_{MSE}( T^M, \phi(S \cdot \overline{\theta(S)}|T\circ \varphi_{TS})) + \mathcal{L}_{MSE}(T^M, \phi(S \cdot \theta(S)|T\circ \varphi_{TS})),
\end{split}
\end{equation}
where $\lambda$ represents the weight that balances the contributions of mutual information loss and similarity loss, and $T^M$ denotes image $T$ after histogram matching. We modify the histogram of $T\circ \varphi_{TS}$ to be similar to that of $S$ in order to mitigate the effects of domain differences.


\section{Experiments}
Our experimental design focuses on two common clinical tasks: atlas-based registration, which involves warping pathology images to a standard atlas template, and longitudinal registration, which involves registering pre-operative images to post-operative images for the purpose of tracking changes over time.

\subsubsection{Dataset and Pre-processing}
For our study, we selected the ICBM 152 Nonlinear Symmetric template as our atlas \cite{ref_MNI152}. We reoriented all MRI scans of the T1 sequence to the RAS orientation with a resolution of 1mm x 1mm x 1mm and align the images to atlas using FreeSurfer \cite{ref_article11}. We then cropped the resulting MRI scans to a size of 160 x 192 x 144, without any image augmentation. To evaluate our approach, we employed a 5-fold cross-validation method and divided our data into training and test sets in an 8:2 ratio.

\noindent{\textbf{\emph{3D brain MRI}}}
OASIS-1 \cite{ref_OASIS} includes 416 cross-sectional MRI scans from individuals aged 18 to 96, with 100 of them diagnosed with mild to moderate Alzheimer's disease. BraTS2020 \cite{ref_BraTS1} provides 369 expert-labeled pre-operative MRI scans of glioblastomas and low-grade gliomas, acquired from multiple institutions for routine clinical use.

\noindent{\textbf{\emph{3D pseudo brain MRI}}}
To evaluate the performance of atlas-based registration, it is essential to have the correct mapping of pathological regions to healthy brain regions. To create such a mapping, we generated a pseudo dataset by utilizing images from the OASIS-1 and BraTS2020. From the resulting t1 sequences, a pseudo dataset of 300 images was randomly selected for further analysis.

\noindent{\textbf{\emph{Real Data with Landmarks}}}
BraTS-Reg 2022 \cite{ref_BraTSReg} provides extensive annotations of landmarks points within both the pre-operative and the follow-up scans that have been generated by clinical experts. A total of 140 images are provided, of which 112 are for training, and 28 for testing. 

\begin{figure}[t!]
\includegraphics[width=\textwidth]{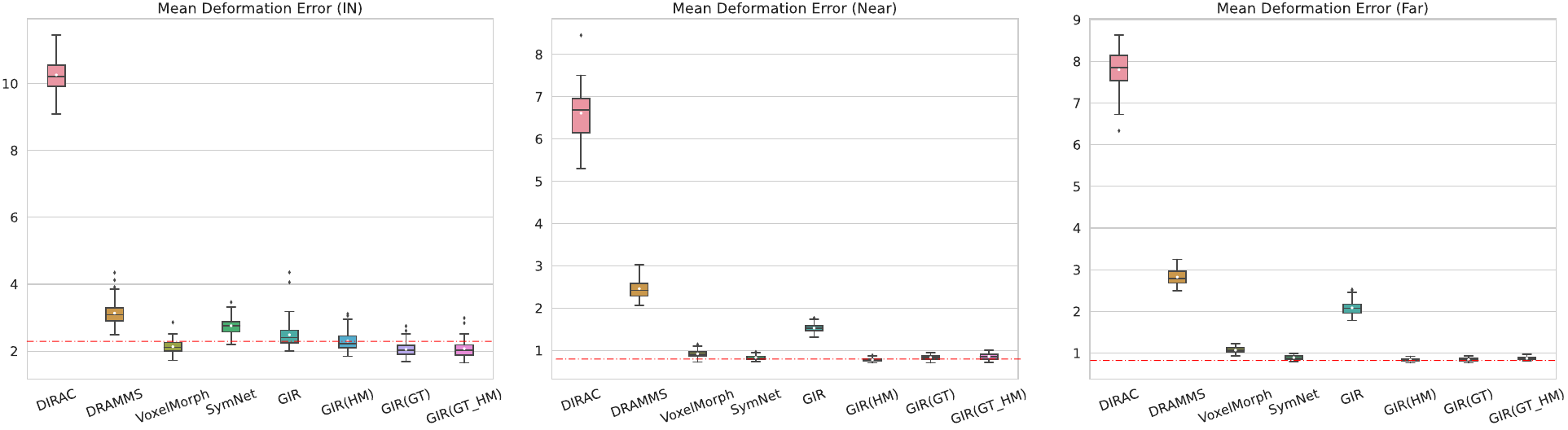}
\caption{Boxplots of mean deformation errors with respect to the gold standard deformations in three different regions on the pseudo dataset. Left to right: in tumor, near tumor and far from tumor.} \label{mre}
\end{figure}

\subsubsection{Comparison to pathology registration}
We compared our method (GIRNet) with competitive algorithms: 1) three cutting-edge deep learning-based unsupervised deformable registration approaches: VoxelMorph \cite{voxelmorph}, VoxelMorph-DF \cite{VM-DF} and SymNet \cite{ref_article31}. 2) two unsupervised deformable registration methods for pathological images: DRAMMS \cite{ref_article14} and DIRAC \cite{DIRAC}. DRAMMS is an optimization-based method that reduces the impact of non-corresponding regions. DIRAC jointly estimates regions with absent correspondence and bidirectional deformation fields and ranked first in the BraTSReg2022 challenge.

\noindent{\textbf{\emph{Atlas-based registration}}}
After creating the pseudo dataset, we warped brain MR images without tumors to the atlas and used the resulting deformation field as the gold standard for evaluation. We then evaluated the mean deformation error (MDE) in three regions: 1) the tumor region. 2) the normal region near the tumor (within 30 voxels). 3) the normal region far from the tumor (over 30 voxels but within brain tissue). Our results, presented in Figure \ref{mre}, show that our method with histogram matching (HM) outperforms other methods in all three regions, particularly in the normal regions (near and far). By utilizing HM, our network achieves an MDE of less than 1 mm compared to the gold standard deformations. 
These results demonstrate the effectiveness of our method in differentiating the impact of pathology in atlas-based registration tasks. Specifically, DIRAC is unable to eliminate the influence of domain differences and resulting in the largest registration error among the evaluated methods.

\begin{figure}[t!]
\includegraphics[width=\textwidth]{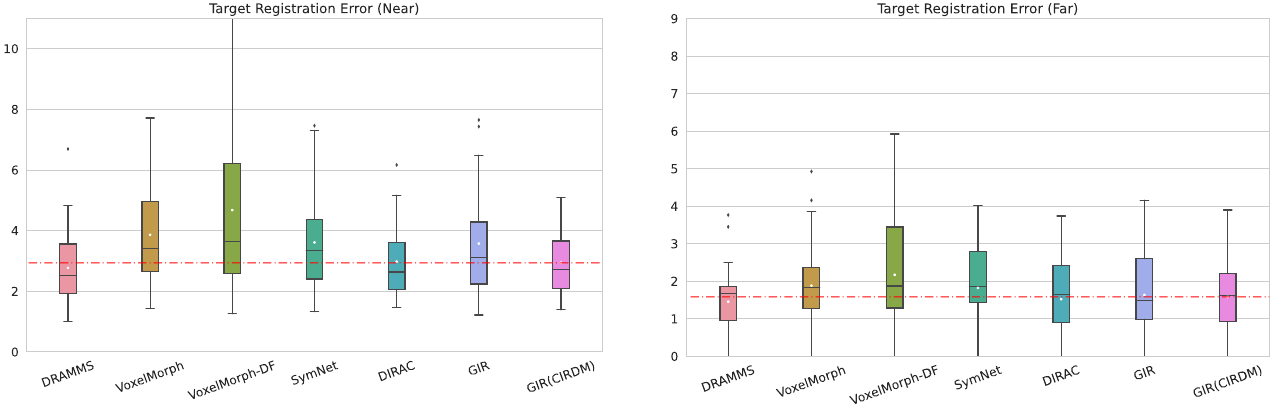}
\caption{Boxplots of the average target registration error (TRE) in two different regions: near tumor (left) and far from tumor (right).} \label{tre}
\end{figure}

\noindent{\textbf{\emph{Longitudinal registration}}}
To perform the longitudinal registration task, we registered each pre-operative scan to the corresponding follow-up scan of the same patient and measured the mean target registration error (TRE) of the paired landmarks using the resulting deformation field. For this purpose, we leveraged SegNet, trained on BraTS2020, to segment the tumor of BraTSReg2022 and separated the landmarks into two regions: near tumor and far from tumor. Figure \ref{tre} shows the mean TRE for the various registration approaches. In our proposed framework, we replaced RegNet  with CIR-DM \cite{CIR-DM} (denoted as GIR(CIRDM)) and achieved comparable performance with the state-of-the-art method DIRAC. Moreover, our GIR approach outperforms other deep learning-based methods and achieved accurate segmentation of pathological images.

\subsubsection{Unsupervised segmentation}
To quantitatively evaluate the segmentation capability of our proposed framework, we compared its performance with other unsupervised segmentation techniques methods, including unsupervised clustering toolbox AUCseg \cite{aucseg}, joint non-correspondence segmentation and registration method NCRNet \cite{ref_article19}, and DIRAC. We used the mean Dice similarity coefficient (DSC) to evaluate the similarity between predicted masks and the ground truth. As shown in Table \ref{tab1}, AUCseg fails to detect the lesion in T1 scans. Our proposed framework achieved the highest DSC result of 0.83, following post-processing.

\begin{table}[!hbtp]
\centering
\caption{Average Dice Similarity Coefficients (DSCs) of Various Model Segmentation Results, Including GIRNet using different techniques: Histogram Matching (HM), Training with ground truth (GT), Mask binarized by threshold 0.5 (TH), and Post-processed by random walker algorithm (PP).}\label{tab1}

\resizebox{\textwidth}{!}{
\begin{tabular}{*{8}c}
\toprule
\multirow{2}*{Dataset} & \multirow{2}*{AUCseg} & \multirow{2}*{NCRNet} & \multirow{2}*{DIRAC} & & \multicolumn{2}{c}{GIRNet} \\
 \cmidrule(lr){5-8}
 &   &  &  & TH & HM+TH & HM+PP & GT \\
\midrule
Pesudo & 0.095(±0.007) & 0.201 & 0.18 & 0.254(±0.03) & 0.744(±0.02) & 0.831(±0.013) &0.921(±0.001)  \\
BraTS2020 & 0.088(±0.010) & 0.191& 0.187 & 0.287(±0.01) & 0.588(±0.014) & 0.611(±0.012) & 0.746(±0.02)  \\
\bottomrule
\end{tabular}
}
\end{table}

\subsubsection{Ablation study}
\begin{figure}[!t]
\includegraphics[width=\textwidth]{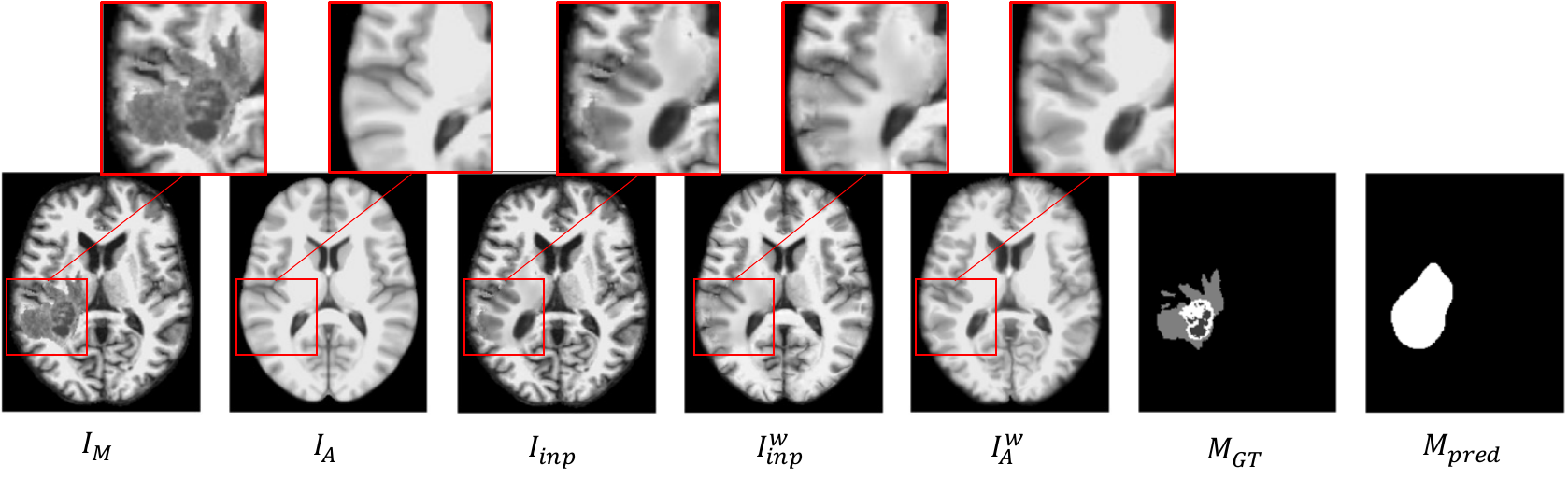}
\caption{Registration and segmentation results for Pseudo dataset. The 7 columns show: 1) the moving image; 2) the atlas; 3) the inpainted image; 4) the warped inpainted image; 5) the warped atlas image; 6) the ground truth mask 7) the predicted mask.} \label{fig4}
\end{figure}

We compared the performance of the InpNet trained with histogram matching (HM) and the SegNet trained with ground truth masks (GT). The results, shown in Table \ref{tab1} and Figure \ref{mre}, demonstrate that domain differences between $S$ and $T$ have a significant effect on segmentation accuracy (without HM), leading to lower registration quality overall. Additionally, Figure \ref{fig4} shows an example of a pseudo image. We reconstructed the spatial correspondence by first using SegNet to localize the lesion and then using InpNet to inpaint it with the normal appearance.
 
\section{Conclusion}
In this paper, we proposed a novel tri-net framework for joint image registration and unsupervised segmentation in medical imaging based on mutual information minimization in collaborative learning. Our experiments demonstrate that the proposed framework is effective for both atlas-based and longitudinal pathology image registration. We also observed that the accuracy of the segmentation network is significantly influenced by the quality of the inpainting, which, in turn, affects the registration outcome. In the future, our research will focus on enhancing the performance of InpNet to address domain differences better to improve the registration results.

\section{Acknowledgements}
This project is supported by NSFC Key Program 62236009,
Shenzhen Fundamental Research Program (General Program) JCYJ 20210324140807019, NSFC General Program 61876032, and Key Laboratory of Data Intelligence and Cognitive Computing, Longhua District, Shenzhen.

\bibliography{paper}

\appendix
\section{Details on Minimal Mutual Information}
\begin{figure}[!hp] 
\includegraphics[width=\textwidth]{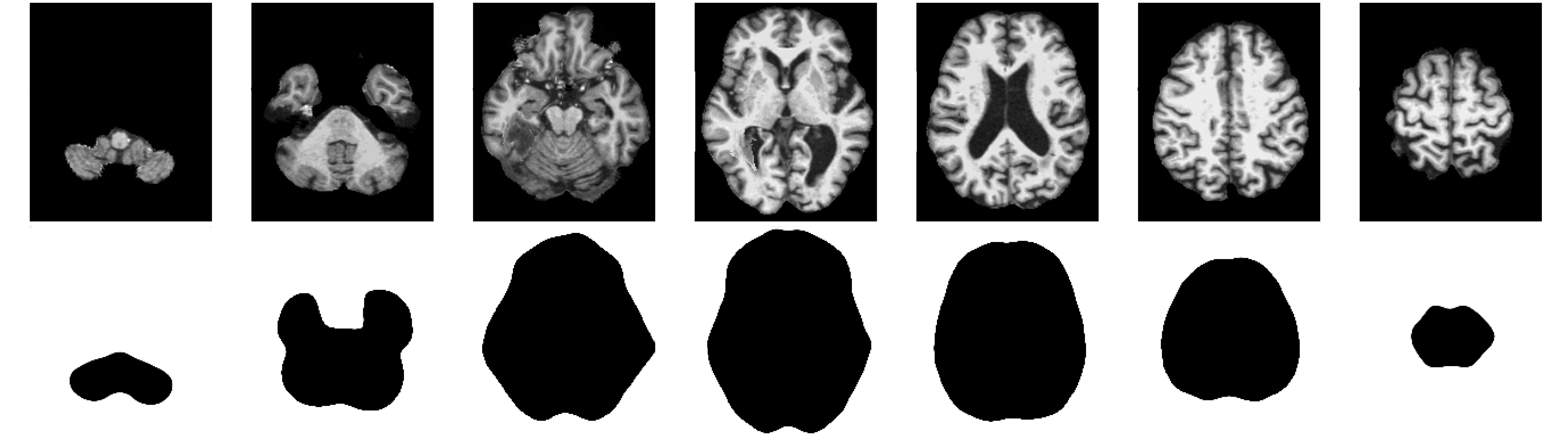}
\caption{Example of MR slices from a pathological brain image, which is an undesirable segmentation of brain tissue and black background based on mutual information theory without semantic information provided by RegNet.} \label{fig1}
\end{figure}

Let $\Omega \in \mathbb{R^d}$ denote the image domain, $F=\Omega \cdot M$ and $B=\Omega \cdot \overline{M}$ denote the foreground and background. The mutual information between foreground and background can be expressed as:
\begin{equation}
I(F,B) = H(F)-H(F|B) = H(B)-H(B|F)
\end{equation}
where $H(\cdot)$ means the (Shannon) entropy, $H(\cdot|\cdot)$ means the conditional entropy. We parameterize the Segmentor as $\theta$, and the purpose of optimization is to find the mask $M$ so that the $F$ and $B$ are independent of each other, i.e., $I(F,B)=0$. To prevent the trivial solution, we use normalized mutual information named \textbf{coefficients of constraint}.


\begin{equation}
\label{eq2}
C(F_{\theta}, B_{\theta}) = \frac{I(F_{\theta},B_{\theta})}{H(F_{\theta})} + \frac{I(B_{\theta},F_{\theta})}{H(F_{\theta})} = 2 - (\frac{H(F_{\theta}|B_{\theta})}{H(F_{\theta})} + \frac{H(B_{\theta}|F_{\theta})}{H(B_{\theta})})
\end{equation}
where $H$ means the (Shannon) entropy. The optimization objective can be simplified as follows:

\begin{equation}
\begin{split}
\label{eq3}
        \mathop{max}\limits_{\theta} \frac{H(F_{\theta}|B_{\theta})}{H(F_{\theta})} + \frac{H(B_{\theta}|F_{\theta})}{H(B_{\theta})} 
        =\mathop{max}\limits_{\theta} \frac{\mathbb{E}_\Omega[logP(\Omega \cdot \theta(\Omega)|\Omega \cdot \overline{\theta(\Omega)})]}{-\mathbb{E}_\Omega[logP(\Omega \cdot \theta(\Omega))]} \\ + \frac{-\mathbb{E}_\Omega[logP(\Omega \cdot \overline{\theta(\Omega)}|\Omega \cdot \theta(\Omega))]}{-\mathbb{E}_\Omega[logP(\Omega \cdot \overline{\theta(\Omega)})]}
\end{split}
\end{equation}
We model the conditional probabilities in the above equation by assuming that they obey a common distribution with identity covariance with a probability density function of $f(x|\mu, I)=\frac{1}{2I}exp(-\frac{ \mathcal{D} (x, \mu)}{I})$, where $\mathcal{D}$ is the distance between the two variables. And We assume the marginal distribution obeys the uniform distribution. Thus, For this probabilistic assumption, we get 
\begin{equation}
\begin{split}
    P(\Omega \cdot \theta(\Omega)|\Omega \cdot \overline{\theta(\Omega)}) &\propto exp(- \mathcal{D}(\Omega \cdot \theta(\Omega), \phi(\Omega \cdot \overline{\theta(\Omega)}))) \\
    P(\Omega \cdot \theta(\Omega)) &\propto exp(- \left \|\theta(\Omega) \right \| )
\end{split}
\end{equation}
where $\phi(\Omega \cdot \overline{\theta(\Omega)})$ is the foreground predicted by inpainter $\phi$, which estimate the conditional means. This setup can be formulated as an adversarial optimization problem, where the SegNet $\theta$ strives to predict the probability of foreground (background) from background (foreground) while maximizing the distance between $F$ and $\phi(B)$. In contrast, the inpainter $\phi$ aims to inpaint the foreground (background) from the background (foreground) while minimizing the distance between $F$ and $\phi(B)$. This can be mathematically represented as:
\begin{equation}
    \begin{split}
       \mathop{max}\limits_{\theta}\mathop{min}\limits_{\phi} \frac{-\mathbb{E}_\Omega[\mathcal{D}(\Omega \cdot \theta(\Omega), \phi(\Omega \cdot \overline{\theta(\Omega)}))]}{-\mathbb{E}_\Omega[\left \| \theta(\Omega) \right \|]} + \frac{-\mathbb{E}_\Omega[\mathcal{D}(\Omega \cdot \overline{\theta(\Omega)},  \phi(\Omega \cdot \theta(\Omega)))]}{-\mathbb{E}_\Omega[\left \| \overline{\theta(\Omega)}\right \|]}
    \end{split}
\end{equation} Figure \ref{fig1} illustrates that accurate segmentation of the lesion cannot be achieved solely through Minimal Mutual Information.

\section{Architecture details}
For simplicity, we use Unet \cite{ref_article30} as the basic structure of SegNet. The encoder consists of 4 convolutional layers and then performs 1/2 downsampling, the decoder part has 7 convolutional layers with 2x upsampling, recovers the feature map to the original image size, the last 3 layers are all convolutions of size 3x3, and there is no normalization in the whole network.

\begin{figure}[!t]
\includegraphics[width=\textwidth]{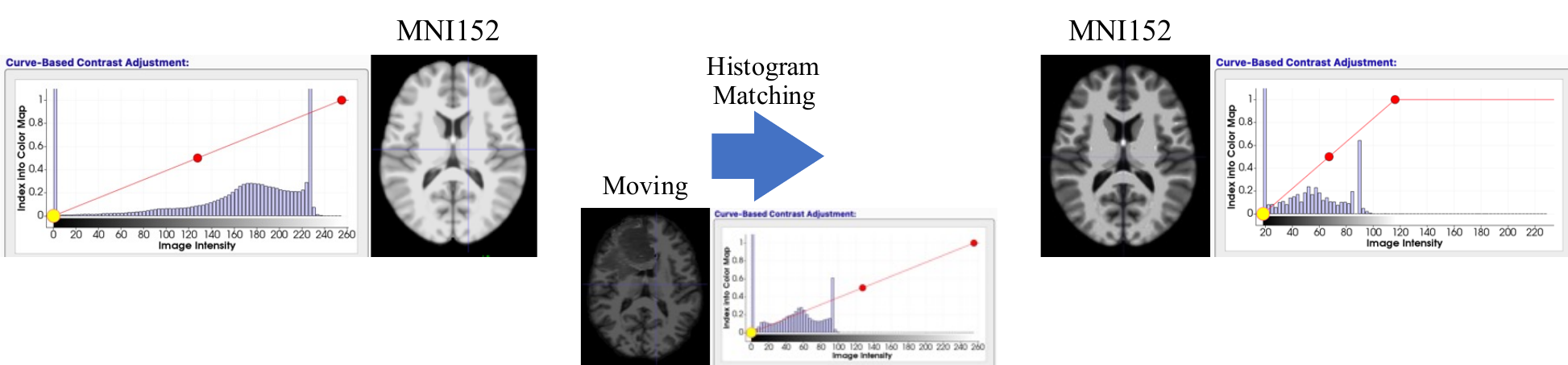}
\caption{One example histogram matching result for MNI152.} \label{fig3}
\end{figure}

\section{Implementation}
All of our code was written on the PyTorch framework and experimented on NVIDIA GeForce TITAN Xp GPUs with 12G of RAM. For training, The initial learning rate is set to 1e-4 and we employ a “poly” decay strategy.  the AdamW optimizer is employed to train the model with a weight decay of 1e-2. The batch size is set to 1, with an epoch of 200. The parameter $\lambda$ of the equation \ref{eq8} is set to 100. In order to speed up the training process, we pretrained the inpainter on OASIS dataset and let it learn to inpaint an image masked by a random size mask by simply copying and pasting.

\begin{figure}[!h]
\includegraphics[width=\textwidth]{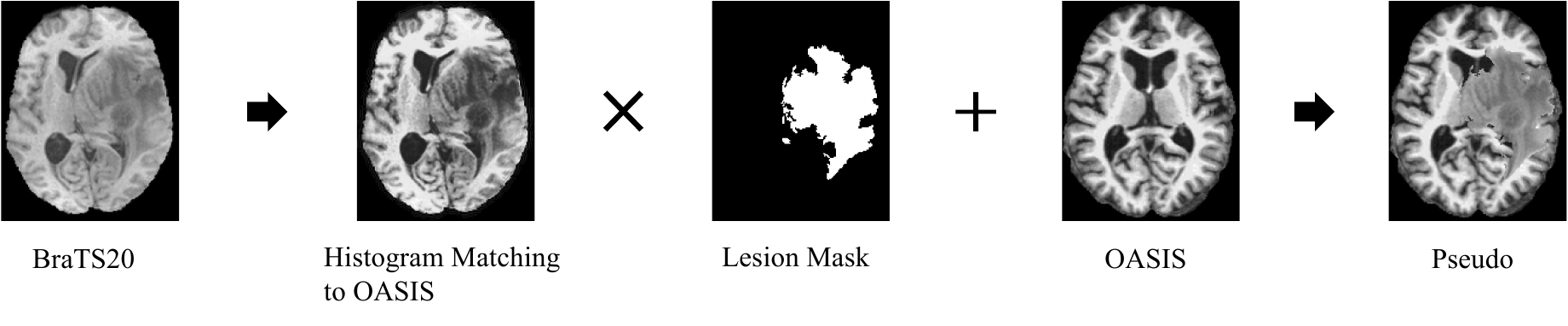}
\caption{We inserted the lesions from BraTS into brain images provided by OASIS by the following steps. 1) Align both OASIS and BraTS MRI to MNI 152. 2) Match intensities of BraTS to intensities of OASIS through histogram matching. 3) Crop the BraTS lesion with the given mask, and then replace the corresponding tissue in OASIS.} \label{pseudo}
\end{figure}

%
%
%
%

\end{document}